# Calibrating conservative and dissipative response of electrically-driven quartz tuning forks

**Lifeng Hao**[1], **Qi Wang**[1], **Ping Peng**[1], **Zhenxing Cao**[1], **Weicheng Jiao**[1], **Fan Yang**[1], **Wenbo Liu**[2], **Rongguo Wang**[1] **and Xiaodong He**[1]

[1]Center for Composite Materials and Structures, Harbin Institute of Technology, Harbin 150080, China

[2]School of Materials Science and Engineering, Harbin Institute of Technology, Harbin 150001, China

E-mail: `hlf@hit.edu.cn`

E-mail: `wrg@hit.edu.cn`

**Abstract.** Determining sensor parameters is a prerequisite for quantitative force measurement. Here we report a direct, high-precision calibration method for quartz tuning fork(TF) sensors that are popular in the field of nanomechanical measurement. In the method, conservative and dissipative forces with controlled amplitudes are applied to one prong of TF directly to mimic the tip-sample interaction, and the responses of the sensor are measured at the same time to extract sensor parameters. The method, for the first time, allows force gradient and damping coefficient which correspond to the conservative and dissipative interactions to be measured simultaneously. The calibration result shows surprisingly that, unlike cantilevers, the frequency shift for TFs depends on both the conservative and dissipative forces, which may be ascribed to the complex dynamics. The effectiveness of the method is testified by force spectrum measurement with a calibrated TF. The method is generic for all kinds of sensors used for non-contact atomic force microscopy(NC-AFM) and is an important improvement for quantitative nanomechanical measurement.

## 1. Introduction

Over the past two decades, quartz tuning forks(TF) have been widely used as force sensors for nanomechanical measurement in many fields, such as atomic force microscopy(AFM), scanning near field optical microscopy(SNOM)[1] and so on[2, 3]. Two working configurations have been normally employed: TF configuration[4], with two prongs oscillate freely; qPlus configuration[5], with the tipless prong glued to a massive substrate. Despite some advantages, such as higher quality factor $Q$, being less vulnerable to vibration of the baseplate[6], TF configuration is less popular for quantitative measurement compared to qPlus configuration due to the difficulty to interpret the measurement results[7]. For qPlus sensors, the force gradient $\Delta k$ can be calculated easily from the frequency shift $\Delta f$, since the qPlus sensor itself is basically a quartz cantilever. However, for TF sensors, the dynamics is complex and what makes it even worse is that the damping mechanism is not well understood[8, 9]. Hence in order to take full advantage of TF sensors for quantitative measurement, a reliable calibration method is required.

Generally speaking, the calibration process is to build the relationship between the detected signals and the relevant physical quantities. For the case of TFs, it is to figure out the dependence of the measured $\Delta f$ and excitation voltage $U$ to the conservative(even) and dissipative(odd) tip-sample interactions which is represented by $\Delta k$ and the damping coefficient $\Gamma$ respectively[10]. In the past, several methods have been borrowed from earlier work on cantilevers to estimate the stiffness of TF, such as Cleveland method[8, 11, 12], noise spectrum method[6, 8, 12], calculation from beam theory[13–15], and finite element simulation method[16]. However, none of them is capable to calibrate the damping effect. In this paper, we report a novel method, in which both the conservative and dissipative responses of TFs can be calibrated in a high-precision and non-invasive manner.

The rest of the paper is organized as follows: we begin with the description of the basic idea behind the method and demonstrate that all possible tip-sample interactions can be generated by an AC force applied to TF in case that the interaction is expressed as the combination of a conservative and a dissipative force. And the AC force should resonate at the same frequency as TF with tunable amplitude and relative phase with respect to the displacement of TF. Next, we described the experiment setup for implementing the calibration, which consists of a specially designed dual-output phase locked loop(PLL) and a laser illumination system. With the method, a TF with a carbon fiber tip glued to one prong is calibrated, and the dependence of $\Delta k$ and $\Gamma$ on $\Delta f$ and $U$ are studied. In order to test the effectiveness of the method, the calibrated sensor is used as force sensor to collect force spectrum on a highly oriented pyrolytic graphite(HOPG) sample, and $\Delta k(z)$ and $\Gamma(z)$ are extracted successfully for each tip-sample displacement. Finally, the effect of the tiny structure differences and added masses on the performance of TFs are investigated by calibrating a series of identical TFs.

## 2. Theory

The key point of the calibration method is to measure the response of TF when applying a force to it. The applied force should cover all possible tip-sample interactions in a controlled manner. In the following we will consider that the TF is excited at or near its natural frequency $f_0$ with a constant oscillation amplitude $A_0$. Let $x = A_0 \cos(2\pi f t)$ be the displacement of the end of the TF prong with mass added. $f$ is the actual resonance frequency and $\Delta f$ is equal to $f - f_0$. The interaction between the tip and the sample can be expressed as

$$F_{\text{int}} = F_{\text{con}} + F_{\text{dis}}, \tag{1}$$

in which $F_{\text{con}}$ is the conservative force and $F_{\text{dis}}$ is the dissipative force. According to [10], $F_{\text{con}}$ and $F_{\text{dis}}$ are in phase with $x$ and the velocity $\dot{x}$ respectively. Hence, $F_{\text{int}}$ can be expressed in a general form as

$$F_{\text{int}} = a \cos(2\pi f t) + b \sin(2\pi f t), \tag{2}$$

in which $a \cos(2\pi f t)$ and $b \sin(2\pi f t)$ corresponds to $F_{\text{con}}$ and $F_{\text{dis}}$ respectively. By varying the value of $a$ and $b$, an arbitrary force can be generated. For the sake of simplicity, we define $F_0 = \sqrt{a^2 + b^2}$ and $\theta = \arctan(a/b)$, then equation(2) can be rewritten as

$$F_{\text{int}} = F_0 \sin(2\pi f t + \theta), \tag{3}$$

which consists of the contributions of both $F_{\text{con}}$ and $F_{\text{dis}}$ that can be expressed as

$$F_{\text{con}} = F_0 \sin\theta \cos(2\pi f t), \tag{4a}$$

$$F_{\text{dis}} = F_0 \cos\theta \sin(2\pi f t). \tag{4b}$$

Equation(3) is equivalent to equation(2) but reveals a clear physical meaning and is practically easy to generate. It shows that an arbitrary force can be mimicked by a force oscillating at the resonance frequency of TF, with a constant amplitude and a fixed phase difference with respect to $x$. Theoretically, all possible combination of $F_{\text{con}}$ and $F_{\text{dis}}$ can be generated by varying $F_0$ and $\theta$. For frequency modulated AFM, the language of $\Delta k$ and $\Gamma$ is always preferred, instead of $F_{\text{con}}$ and $F_{\text{dis}}$. $\Delta k$ can be calculated by [17]

$$\Delta k = \frac{2f}{A_0} \int_0^{\frac{1}{f}} F_{\text{con}} \cos(2\pi f t) dt, \tag{5}$$

substituting equation(4a) into equation(5) gives

$$\Delta k = \frac{F_0}{A_0} \sin\theta. \tag{6}$$

According to [10]

$$F_{\text{dis}} = \Gamma \dot{x}, \tag{7}$$

substituting equation(7) into equation(4b),we get

$$\Gamma = -\frac{F_0}{2\pi f A_0} \cos\theta. \tag{8}$$

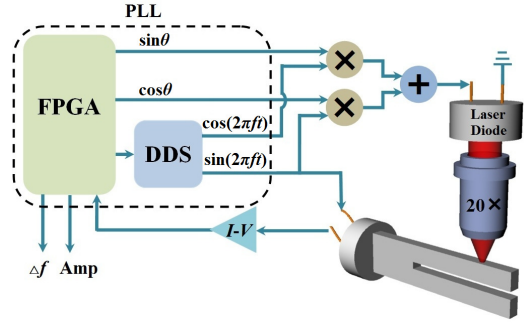

**Figure 1.** Block diagram of the experimental setup.

It is clear, from equation (6) and (8), that we can tune the magnitudes of $\Delta k$ and $\Gamma$ by either $F_0$ or $A_0$, and their ratio by $\theta$.

## 3. Experimental setup

Figure 1 shows the block diagram of our experimental setup to generate the force as expressed by equation(3). The setup mainly consists of a dual-output PLL and a laser illumination system.

PLL is the most common method to detect the resonance frequency of TF. Various home-build and commercial versions are available[18–20]. The circuit presented in figure 1 is a modified version of the hybrid analogy digital PLL reported earlier[17], which consists of two parts: the basic circuit to provide normal PLL functions and the extra one to generate the phase shifted output. The basic PLL circuit is composed of a four quadrant multiplier (AD734) as phase deflector, a second order analog butterworth low pass filter plus a digital proportional-integral(PI) feedback as loop filter and a 48-bit direct digital synthesis (DDS, AD9854) as a voltage controlled oscillator. The butterworth filter is constructed from two pieces of precision high speed op-amp(OPA627) and the digital PI feedback is implemented in a field programmable gate array (FPGA, Altera CycloneIV EP4CE15). Besides, another digital PI feedback is also encoded in FPGA to keep the oscillation amplitude at a set value. During operation, the current generated from the oscillation of TF is converted into a voltage signal via a stray capacitance compensated transimpedance amplifier[21], and consecutively fed into PLL. Phase and amplitude of the input signal are extracted and fed into the frequency and amplitude PI feedbacks separately. The output of the frequency PI feedback is used to control DDS, which generate reference signal $\cos(2\pi ft)$ and $\sin(2\pi ft)$, as shown in figure 1. Among them, $\sin(2\pi ft)$ multiplied by the output of the amplitude PI feedback (not shown in figure 1), also works as the first output to drive TF sensors.

The aforementioned circuit only provides the basic function of a normal PLL. Based on it, an extra circuit is designed to generate the second phase-shifted output. As shown in figure 1, a phase signal $\theta$ is generated in FPGA. Then $\sin\theta$ and $\cos\theta$ are calculated and output via digital-to-analog converter (DAC8811) and multiplied with the reference signal $\cos(2\pi ft)$ and $\sin(2\pi ft)$ respectively, giving $\sin\theta\cos(2\pi ft)$ and $\cos\theta\sin(2\pi ft)$. The two signals are added together resulting in $\sin(2\pi ft+\theta)$. By multiplying it with a voltage $U_a$ generated by FPGA as well, we then get the second output from PLL, which is phase shifted from $\sin(2\pi ft)$, the driving signal of TF, by $\theta$ and by $\theta+\pi/2$ from $x$‡. It is noteworthy that both $U_a$ and $\theta$ are generated from FPGA and can be accurately adjusted. The tuning range for $U_a$ is [0, 10V] while that for $\theta$ is [0, $2\pi$]. Hence the second output $U_a\sin(2\pi ft+\theta)$ satisfied the requirement of equation(3) and is capable to mimic the arbitrary tip-sample interactions. The phase difference between the 1st and 2nd output are calibrated by a commercial lock-in amplifier SRS785, since several analog components are used, which may introduce some phase errors.

The second output is fed into a laser diode (5mW at 520nm) to emit a modulated laser beam. The beam, after focused by a 20×objective lens, is used to excite TF sidewards via photothermal effect[22, 23], as shown in figure 1. Hence, the voltage signal from the second output of PLL is converted into an oscillating force applied to TF with the phase shifted by $\theta+\pi/2$ with respect to $x$.

## 4. Result and discussion

### *4.1. Calibration of a tuning fork sensor*

A TF sensor (DT206, KDS Daishinku Corp.) with a carbon fiber tip glued on one prong was calibrated by the setup mentioned above. The carbon fiber tip was electrochemically etched in 4M NaOH solution[24]. In order to get a controlled stimulus to TF, $F_0$ should be calibrated first. $F_0$ is generated by laser illumination and its magnitude is determined by the illumination intensity. It is noteworthy that several factors affect the magnitude of $F_0$, including the driving voltage, the distance between the TF and the laser diode, the laser spot size and illumination position on TF. Hence, it is practically difficult to deduce $F_0$ from the illumination intensity directly. Here we followed a different approach using the equivalence among different driving methods to acquire the value of $F_0$. During operation, TF is basically an oscillator resonating at its natural frequency with energy dissipation compensated by extra excitation. The energy dissipated is characterized by $Q$ which is defined as $Q=2\pi\frac{\Delta E}{E}$, in which $E$ is the energy store per cycle, equal to $\frac{1}{2}k_\text{eff}A^2$ for TF and $\Delta E$ is the energy dissipated per cycle. $Q$ is intrinsic quantity for TF, independent of the way it is excited. Therefore, $\Delta E$ is the same for different excitation methods in case that the oscillation amplitudes are the same. So it is possible to use this equivalent relationship to determine an unknown excitation from a known one. Here we use the electrical excitation to calibration the photothermal excitation. We implement the laser illumination as driving signal to excite the TF to measure $f_0$ and $Q$ and compare them with those measured from electrical excitation as shown in table 1. We note that $f_0$ are the same while $Q$ are very close as expected. For electrical excitation,

$$\Delta E_{\text{ele}} = \int_0^{\frac{1}{f_0}} UI dt = \frac{U_0 I_0}{2\pi f_0}, \qquad (9)$$

where $U = U_0 \sin(2\pi f_0 t)$ and $I = I_0 \sin(2\pi f_0 t)$ are the voltage and current between the two electrodes of the TF sensor. For photothermal excitation, the modulated illumination is equivalent to an AC force $F_{\text{photo}} = F_0 \sin(2\pi f_0 t)$ applied to the prongs of the TF sensor, and

$$\Delta E_{\text{photo}} = \int_0^{\frac{1}{f_0}} F_{\text{photo}} dx = F_0 A_{\text{photo}}, \qquad (10)$$

in which $x_{\text{photo}} = A_{\text{photo}} \cos(2\pi f_0 t)$. If we make the oscillation amplitude for electrical excitation equal to that for photothermal excitation, we have

$$\frac{\Delta E_{\text{photo}}}{Q_{\text{photo}}} = \frac{\Delta E_{\text{ele}}}{Q_{\text{ele}}}, \qquad (11)$$

‡ On resonance, the driving signal of TF is always 90° out of phase with respect to the displacement for TF.

Table 1. Resonance frequency and quality factor for the calibrated TF sensor under electrical and photothermal excitations.

| Excitation | $f_0$(Hz) | Q |
|---|---|---|
| Electrical | 32 265 | 8248 |
| Photothermal | 32 265 | 8279 |

Hence, the amplitude of the oscillating force can be expressed as

$$F_0 = \frac{U_0 I_0}{2\pi f_0 A_{\text{photo}}} \cdot \frac{Q_{\text{photo}}}{Q_{\text{ele}}}. \qquad (12)$$

$U_0$ and $f_0$ come from the driving signal with known magnitude. $I_0$ can be measured by the pre-amplifier accurately. $A_{\text{photo}}$ is measured by optical interference[25] and can be accurate to better than 1% with careful operations. For the TF sensor, we have $A_{\text{photo}} = 2.99$nm, $U_0 = 1.15$mV and $I_0 = 5.79$nA, giving $F_0 = 10.89$nN. All the parameters in equation(12) can be measured with high precision, so $\Delta k$ and $\Gamma$ applied to the TF sensor can be calculated accurately from equation(6) and (8), and the accuracy of the calibration method is thus guaranteed. It is noteworthy that $F_0$ should be kept constant for the rest of the calibration process which means that both the TF and the laser diode should not be moved and the driving voltage on the diode should be set at a fixed value.

With $F_0$ acquired above, the TF can be calibrated. It is controlled to oscillate at its resonance frequency in a constant oscillation amplitude mode by the basic PLL circuit mentioned above, $\Delta k$ and $\Gamma$ are then applied by laser illumination. For a fixed $A_0$, $\theta$ is ramped from 0 to $2\pi$, giving continually varied $\Delta k$ and $\Gamma$ whose values are calculated by equation(6) and (8), and shown in figures 2(a). At the same time, $\Delta f$ and the amplitude of the excitation voltage $U$ are measured and their dependance on $\theta$ are shown in figure 2(b) and (c). The aforementioned process is repeated for several times for a series of $A_0$ (6.3nm, 7.0nm, 7.7nm, 8.4nm, 9.1nm, 9.8nm, 10.5nm). It is noteworthy that $\theta$ varies the ratio between $\Delta k$ and $\Gamma$, while $A_0$ modifies their amplitudes. Hence, by tuning $\theta$ and $A_0$, it is possible to generate all possible combinations of $\Delta k$ and $\Gamma$ encountered in the following measurement.

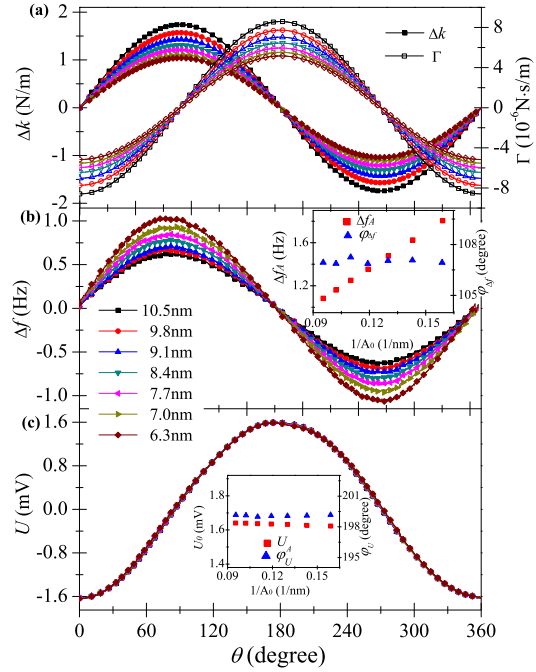

**Figure 2.** (Color online)Calibration results: applied force gradient and damping coefficient to the prong of the TF (a), measured $\Delta f$ (b) and $U$ (c) as a function of the relative phase shift between the two outputs of dual-output PLL. $\Delta f$ and $U$ curves are fitted to sine functions and the fitted amplitudes and phase shifts are plotted versus the inverse of $A_0$ for $\Delta f$ and $U$ in the insets of (b) and (c) respectively.

For a certain $A_0$, both $\Delta f$ and $U$ show a sine-like behavior and can be fitted as

$$\Delta f(\theta) = \Delta f_A \sin(\theta - \varphi_{\Delta f}), \qquad (13a)$$
$$U(\theta) = U_A \sin(\theta - \varphi_U), \qquad (13b)$$

in which $\Delta f_A$ and $U_A$ are the amplitudes of the sine functions and $\varphi_{\Delta f}$ and $\varphi_U$ are the phase shifts. It is found that $\Delta f_A$ is linearly dependent on $1/A_0$ as shown in the inset of figure 2(b), and $U_A$ is roughly constant as shown in the inset of figure 2(c). This is reasonable since $\Delta f$ is supposed to be proportional to $\Delta k$ which is inverse to $A_0$, and $U_A$ is proportional to $F_0$ which is kept constant in the experiment. Besides, both $\varphi_{\Delta f}$ and $\varphi_U$ are constant with the values of $-4.1 \pm 0.2°$ and $88.0 \pm 0.1°$ respectively. Hence, equation(13) can be rewritten as

$$\begin{pmatrix} \Delta f \\ \frac{U}{A_0} \end{pmatrix} = \begin{pmatrix} a_{11} & a_{12} \\ a_{21} & a_{22} \end{pmatrix} \cdot \begin{pmatrix} \Delta k \\ 2\pi f \Gamma \end{pmatrix}, \qquad (14)$$

in which

$$\begin{pmatrix} a_{11} & a_{12} \\ a_{21} & a_{22} \end{pmatrix} = \begin{pmatrix} \frac{\Delta f_A A_0 \cos \varphi_{\Delta f}}{F_0} & \frac{\Delta f_A A_0 \sin \varphi_{\Delta f}}{F_0} \\ \frac{U_A \cos \varphi_{U_0}}{F_0} & \frac{U_A \sin \varphi_{U_0}}{F_0} \end{pmatrix}. \quad (15)$$

We note that, for a certain TF sensor, $a_{11}$, $a_{12}$, $a_{21}$ and $a_{22}$ are all constants with values of 0.5880Hz·nm/nN, -0.0423Hz·nm/nN, 0.0053mV/nN and 0.1488mV/nN respectively §. Among them, $a_{11}$ is constant showing that $\Delta f$ is proportional to $\Delta k$, in agreement with previous reports in literature, and the effective stiffness of the TF sensors $k_{\text{eff}}$ can be calculated by $k_{\text{eff}} = f_0/2a_{11}$, if assuming $\Delta f/f_0 = \Delta k/2k_{\text{eff}}$. For the sensors in calibration $k_{\text{eff}}$ is equal to 27.44kN/m. $a_{22}$ corresponds to the dependence of $U$ on $\Gamma$, which has not been calibrated yet. Our observation that $a_{22}$ is constant confirms that the amplitude of TF is proportional to the damping coefficient, the same as normally encountered for a cantilever. Both $a_{11}$ and $a_{22}$ behave as expected, however the existence of non-zero $a_{12}$ and $a_{21}$ lead to some surprising results. $a_{21}$ shows the influence of $\Delta k$ on $U$, revealing an interesting fact that conservative interaction can also cause damping. However, this is consistent to earlier observation, and is attributed to the imbalanced oscillation of TF[9, 26]. $a_{12}$ shows the dependance of $\Delta f$ on $\Gamma$, and it is negative showing that damping on TF sensors will cause negative frequency shift. This has never been taken into consideration before. However, this point can be confirmed by a fact, usually observed but always ignored, that every time when a TF is taken out of its canister, its resonance frequency tends to drop for several Hertz. For the TF used here, $f_0$ dropped from 32 768.0Hz to 32 745.2Hz, as list in table 2. The only difference for a TF inside or outside a canister is the increase of the damping coefficient due to the change of the pressure. Hence the frequency drop can only be explained by the existence of a negative $a_{21}$. From the discussion above, it is clear that TF sensors behave differently compared to cantilevers for which $\Delta f$ is solely determined by $\Delta k$ and $U$ by $\Gamma$. For TF sensors, $\Delta f$ comes mainly from $\Delta k$ but with a small correction from $\Gamma$ which is represented by $a_{12}$, and likewise similar correction represented by $a_{21}$ exists for $U$. $a_{12}$ and $a_{21}$ are necessary corrections for quantitative measurement with TF sensors, but are not accounted for in the calibration methods report previously. Actually these two parameters can only be determined by the direct calibration method proposed here with both $F_{\text{con}}$ and $F_{\text{dis}}$ being taken into consideration. For the sake of simplicity, equation(14) can be rewritten as

$$\begin{pmatrix} \Delta k \\ 2\pi f \Gamma \end{pmatrix} = \begin{pmatrix} a_{11} & a_{12} \\ a_{21} & a_{22} \end{pmatrix}^{-1} \cdot \begin{pmatrix} \Delta f \\ \frac{U}{A_0} \end{pmatrix}. \quad (16)$$

Hence, each measured data point $(\Delta f(z), U(z))$ can be converted into $(\Delta k(z), \Gamma(z))$.

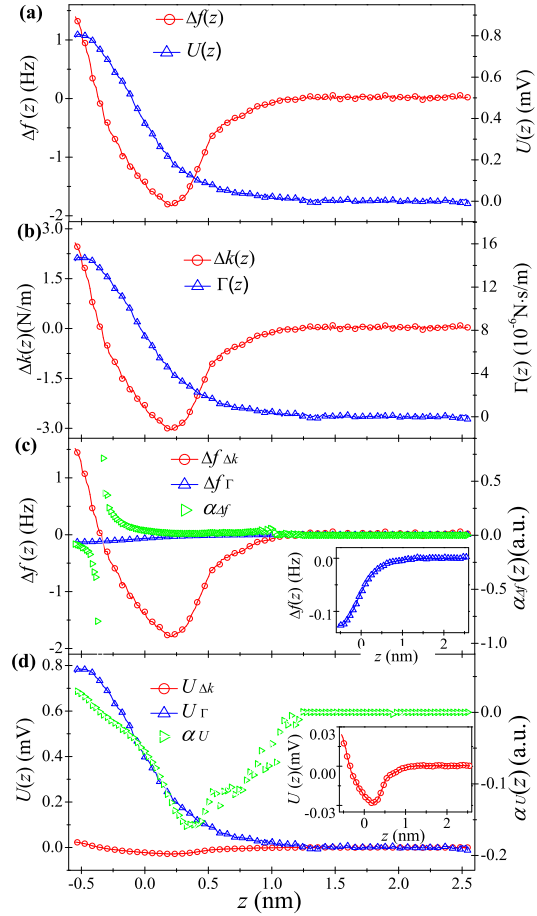

**Figure 3.** (Color online) Quantitative force measurement results: (a) measured $\Delta f(z)$ and $U(z)$ as a function of tip sample displacement; (b) calculated $\Delta k(z)$ and $\Gamma(z)$ curves; (c) frequency shift generated by $\Delta k(z)$ and $\Gamma(z)$ and their ratio; (d) oscillation amplitude generated by $\Delta k(z)$ and $\Gamma(z)$ and their ratio. For the measurement, $f_0$ is 32 265Hz, $A_0$ is 1.77nm and the bandwidth of PLL is 50Hz.

### 4.2. Quantitative force measurement

Quantitative force measurement was carried out with the calibrated TF sensor to test its performance and the validity of the method. We used a custom-made AFM system operated in an inert gas protected environment and controlled by open source SPM software package GXSM and its supported controller(model MK3-PLL, form softdB inc.)[27]. The preamplifier is located very close the sensor and connected to it via a pair of 5cm long bare wires to eliminate the noise generated from cables. Figure 3(a) shows the force spectrum collected on a freshly cleaved HOPG sample. Each data point ($\Delta f(z)$, $U(z)$) on the curves corresponds to an interaction state experienced by TF sensor and, by putting it into equation(16), ($\Delta k(z)$, $\Gamma(z)$) can be calculated as shown in figure 3(b). Hence, both the conservative and dissipative tip-sample interactions are easily extracted from the measurement result. Especially, the determination of $\Gamma$ makes it possible for TF sensors to be used for quantitative investigation of dissipation process in nanoscale[28]. It is also noted that $\Delta k(z)$ and $\Gamma(z)$ curves resemble $\Delta f(z)$ and $U(z)$ curves respectively as expected due to the fact that $a_{11} \gg a_{12}$ and $a_{22} \gg a_{21}$.

§ $F_0$ are set at a fix value. $\varphi_{\Delta f}$, $\varphi_{U_0}$ are constants from the measurement result. And $\Delta f_A \cdot A_0$ are also constant since $\Delta f_A$ is inversely proportional to $A_0$. Hence, $a_{11}$, $a_{12}$, $a_{21}$ and $a_{22}$ are all constants.

To quantify the effects of $a_{12}$ and $a_{21}$, we calculated $\Delta f_{\Delta k}(z)$ and $\Delta f_\Gamma(z)$, the frequency shift caused by $\Delta k(z)$ and $\Gamma(z)$, and the ratio of them, which is expressed as

$$\alpha_{\Delta f}(z) = \frac{\Delta f_\Gamma(z)}{\Delta f_{\Delta k}(z)}, \tag{17}$$

and shown in figure 3(c). The same work is done for $U(z)$ with $U_{\Delta k}(z)$, $U_\Gamma(z)$ and $\alpha_U(z)$ representing the variation of $U(z)$ caused by $\Delta k(z)$ and $\Gamma(z)$ and their ratio, as shown in figure 3(d). Here $\alpha_U(z)$ is defined as

$$\alpha_U(z) = \frac{U_{\Delta k}(z)}{U_\Gamma(z)}. \tag{18}$$

We note that $\alpha_{\Delta f}(z)$ and $\alpha_U(z)$ depend on $z$ in a different manner: $\alpha_{\Delta f}(z)$ is predominant near the point where $\Delta f(z)$ turns from negative to positive; $\alpha_U(z)$ shows a valley in the attractive force region. The maximum value for $\alpha_{\Delta f}(z)$ and $\alpha_U(z)$ are 0.723 and -0.163 respectively, indicating that although the absolute values of $\Delta f_\Gamma(z)$ and $U_{\Delta k}(z)$ are small, they play an important role in certain parts of the force spectrum and should not be simply ignored in quantitative force measurement.

*4.3. Performance differences among identical TF sensors*

Most of TF sensors are manually assembled, so each of them may possess different character due to the tiny structure differences and the variation of added mass from case to case[29]. Hence it is of great importance to study the performance variation for different sensors and also their dependence on mass addition. Here we choose a series of identical TF sensors as mentioned above and glued different masses to some of them. Each TF is calibrated by the proposed method, and their properties are listed in table 2. The first three TFs (S1-3) have no mass glued, and their $f_0$ are the same (32 754.2Hz) and their $Q$ are all about 10 000. As crystal oscillators, they can be regarded as identical, since they can oscillate exactly at the desired frequency. However as force sensors, their performance are quite different, for example $k_\text{eff}$ of S2 is 28% larger than that of S1, and $a_{21}$ of S3 is almost 32 times larger than that of S1. The result clearly reveals the influence of the tiny structure difference on the behavior of TFs. Besides, TF S4-S10 have different masses added to one prong, and it is noted that, for all of them, the values of $Q$, $f_0$, $k_\text{eff}$, $a_{11}$, $a_{12}$, $a_{21}$ and $a_{22}$ are scattered without direct correspondence. This further confirms that structure difference and added mass will strongly influence the performance of TF sensors, suggesting that each TF sensor should be carefully calibrated before being used in quantitative nanomechanical measurement.

Table 2. $Q$, $f_0$, $k_{\text{eff}}$, $a_{11}$, $a_{12}$, $a_{21}$ and $a_{22}$ for a series of identical TFs with different masses added.

| No. | $Q$ | $f_0$ (Hz) | $k_{\text{eff}}$ (kN/m) | $a_{11}$ (Hz·nm/nN) | $a_{12}$ (Hz·nm/nN) | $a_{21}$ (mV/nN) | $a_{22}$ (mV/nN) |
| --- | --- | --- | --- | --- | --- | --- | --- |
| S1[a]  | 10 258 | 32 754.2 | 22.02 | 0.7437 | -0.0522 | -0.0006 | 0.1593 |
| S2[a]  | 10 231 | 32 754.2 | 30.64 | 0.5337 | -0.0278 | 0.0119 | 0.1345 |
| S3[a]  | 10 218 | 32 754.2 | 23.45 | 0.6988 | -0.0226 | -0.0193 | 0.1496 |
| S4[b]  | 8248 | 32 265.0 | 27.44 | 0.5880 | -0.0423 | 0.0053 | 0.1488 |
| S5 | 8878 | 32 248.3 | 28.00 | 0.5759 | -0.0091 | -0.0166 | 0.1499 |
| S6 | 6587 | 31 286.5 | 22.69 | 0.6895 | -0.0337 | -0.0154 | 0.1723 |
| S7 | 6585 | 31 278.2 | 22.43 | 0.6972 | -0.0155 | 0.0096 | 0.1599 |
| S8 | 5695 | 31 826.0 | 22.54 | 0.7060 | -0.0175 | -0.0082 | 0.1709 |
| S9 | 4788 | 30 768.7 | 26.60 | 0.5788 | -0.0232 | -0.0070 | 0.1577 |
| S10 | 3804 | 31 883.0 | 29.48 | 0.5408 | -0.0213 | -0.0063 | 0.1486 |

[a] Bare TF without mass added.
[b] The TF with carbon fiber tip glued on one prong.

## 5. Conclusion

In summary, we have presented a new calibration method for TF sensors and the setup to implement it. The method possess several advantages: first, conservative and dissipative forces can be calibrated simultaneously, allowing TF to explore damping process easily; second, all the parameters involved can be measured experimentally with high accuracy, giving the method the highest precision among the currently available methods; third, a direct relationship can be built between $\Delta k$, $\Gamma$ and the measured $\Delta f$ and $U$, independent of the geometry of sensors, allowing the method to be easily extended to all kinds of NC-AFM sensors, such as cantilever and so on; last but not least, the method is non-invasive, avoiding contamination or degradation of the sensor. By calibrating a series of TFs, we note surprisingly that, unlike cantilevers, the response of TF sensors to conservative and dissipative forces are mixed, that is, $\Delta f$ and $U$ are both affected by $\Delta k$ and $\Gamma$. This indicates that for quantitative force measurement with TF, the conservative and dissipative interactions must be taken into account together. Beside, we also found that each TF shows unique characters and hence should be calibrated individually.

## Acknowledgments

The works were financially supported by the fund of the Natural Science Foundation of China (No.51303039) and Ph.D. Programs Foundation of Ministry of Education of China (No.20122302120034).